\documentclass{cernrep} 
\usepackage{texnames}
\usepackage[T1]{fontenc}
\usepackage{eurosym}
\usepackage{hyperref}
\hypersetup{
    colorlinks=true,
    citecolor=black,
    linkcolor=black,
    filecolor=black,      
    urlcolor=blue,
}

\begin{document}
\title{Diagnostics Examples from lepton-linacs and FELs}
 
\author {A. Cianchi}

\institute{University of Rome "Tor Vergata"}

\begin{abstract}
There is a big gap between the charming principles ideas and the real implementation of a diagnostic.
In this lecture we review some details that make the difference between a good and bad measurement, highlighting also the relation between the measured quantities and the real ones.

\end{abstract}


\maketitle % this produces the title block
 
\section{Introduction}
 
In this lecture we review selected diagnostics techniques developed for the measurement of the longitudinal and transverse phase space in lepton-linacs and FELs. 
The common points between these kind of machines is the charge single passage feature and the high brightness of these beams. 
We focus not only on the general principles but also on the implementation of these devices. 
This lecture is a focus of few diagnostics looking from the prospective of the implementation of the device and helping to understand the meaning of what is measured and its relation with the real value of the parameter that we are looking for.

\section{R.M.S. emittance and its meaning}

Every particle of a beam can be represented by a point in a transverse phase space. 
All of these points are spread out in a 6D phase space ($x$, $p_{x}$, $y$, $p_{y}$, $z$, $p_{z}$ ). 
For simplicity we imagine that all the points are enclosed in a hyperellipsoid and we call emittance the volume of this. 
The projections of this volume in every transverse and in the longitudinal plane are ellipses.
This schematisation allows us to easily describes the beam transport and we call emittance (horizontal, vertical, longitudinal) the area of the ellipse in that phase space. 
Remember that phase space diagrams have momentum (for instant p$_{x}$) on the $y$-axis, while the trace space diagram has the transverse velocity $x^{\prime}$.
The area in the phase space is the Liouville invariant emittance, while the area in the trace space is just the geometrical emittance. 
This is the quantity that we measure, while the normalized emittance is the value that we are interested on.

\begin{figure}[h]
  \centering
  \includegraphics[width=80mm]{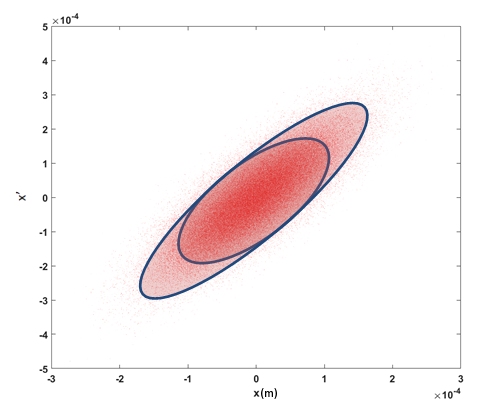}
  \caption{A simple representation of particles in trace space. Two different ellipse are drawn. Which one can be considered to be descriptive of the beam?}
  \label{tracespace1}
\end{figure}

Which is the ellipse that we have to consider to be representative of the beam?
In Fig.\ref{tracespace1} are shown two different ellipses. 
Which one is the best suited for our purpose?

\begin{figure}[h]
  \centering
  \includegraphics[width=80mm]{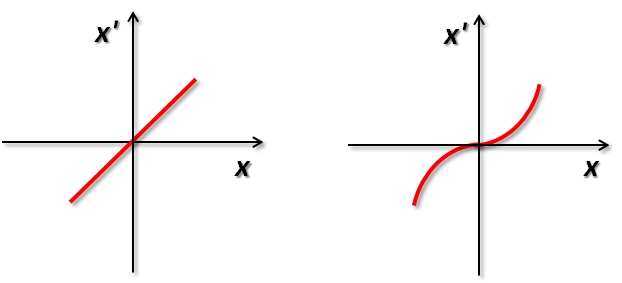}
  \caption{Example of two different trace spaces where there is a linear and quadratic correlation respectively between $x$ and $x^{\prime}$. Both figures have null area.  As it explained in the text only the first one has zero r.m.s. emittance.}
  \label{tracespace2}
\end{figure}

To avoid such complicated and arbitrary decision and because the beam does not have sharp edges, the r.m.s. (root mean square) emittance is used\cite{bib:buon}. 
This is a very well known concept and we suggest the reader to refer to bibliography about that.
The same concept of an area of the phase space is lost in the case of r.m.s. emittance, as it is shown in Fig.\ref{tracespace2}.

In both plots of \Fref{tracespace2} the area is zero, while the r.m.s. emittance is not always zero.
The square of geometrical r.m.s. emittance is defined as
\begin{equation}
\varepsilon ^2 _{\mathrm{rms}}  = \left\langle {x^2 } \right\rangle \left\langle {x'^2 } \right\rangle  - \left\langle {xx'} \right\rangle ^2 \ ,
\label{formula3}
\end{equation}
and assuming a correlation between $x$ and $x^{\prime}$ such as $x^{\prime}=Cx^{n}$, where $C$ is a constant, the geometrical emittance can be rewritten as
\begin{equation}
\varepsilon ^2 _{\mathrm{rms}}  =C^{2}\left\langle {x^2 } \right\rangle \left\langle {x^{2n} } \right\rangle  -C^{2} \left\langle {x^{n + 1} } \right\rangle ^2 \ .
\label{formula4}
\end{equation}
It is clear that if $n=1$, the r.m.s. emittance is zero. 
Looking at Fig.\ref{tracespace2}, being the area zero, this value is consistent with the definition of the emittance as the area.
But for $n>1$ the area is still zero, but the r.m.s. emittance is not.
So the r.m.s. emittance is more a figure of merit of the beam quality than the area of the trace space.
For pure monochromatic beams, the normalized emittance is $\beta\gamma$ times the geometrical emittance, where $\gamma$ is the relativistic factor and $\beta$ is the ratio between the particle speed and the speed of the light.
Otherwise (see, for instance, Refs. \cite{bib:klaus} and \cite{bib:migliorati}), there is also a contribution from the energy spread and the beam divergence.
In fact, the Liouville invariant is the normalized emittance, defined as

\begin{equation}
\varepsilon _\mathrm{n}^2  = \left\langle {x^2 } \right\rangle \left\langle {\beta ^2 \gamma ^2 x'^2 } \right\rangle  - \left\langle {x\beta \gamma x'} \right\rangle  \ .
\label{normalizedemittance}
\end{equation}
It has been demonstrated in \cite{bib:migliorati} that this formula can be rewritten as

\begin{equation}
\varepsilon _\mathrm{n}^2  = \left\langle \gamma  \right\rangle ^2 \sigma _\varepsilon ^2 \left\langle {x^2 } \right\rangle \left\langle {x'^2 } \right\rangle  + \left\langle {\beta \gamma } \right\rangle ^2 \left( {\left\langle {x^2 } \right\rangle \left\langle {x'^2 } \right\rangle  - \left\langle {xx'} \right\rangle ^2 } \right) \ ,
\label{normemittance}
\end{equation}
where $\sigma _\varepsilon$ is the relative energy spread and the second term is the geometrical emittance multiplied by $\beta\gamma$.
The first term is usually negligible in conventional accelerators, but it could be the leading one in some scenarios, such as in plasma-based accelerators.
In such a case, both energy spread and angular divergence are needed, to measure the normalized emittance.

\section{Measuring transverse emittance}

\begin{figure}[ht]
  \centering
  \includegraphics[width=50mm]{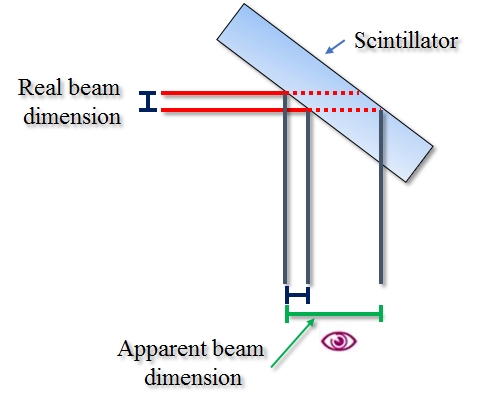}
  \caption{Blurring due to transparent scintillator. The effect is highly exaggerated. The radiation is emitted all along the material, resulting in an apparent increase of the beam size.}
  \label{blurring}
\end{figure}

To measure the r.m.s. transverse emittance, whatever is the method, we need to measure the r.m.s beam transverse size. 
In a linac intercepting diagnostics are allowed, except in the cases where too much power is deposited in these devices and it can seriously damage them. 
If it is not the case, a scintillator screen \cite{bib:kube} is usually the best choice, because it allows to make a 2D picture of the beam. 
The light emission in a scintillator is a bulk effect, i.e. the light is emitted all along the path of the particle. 
In Fig.\ref{blurring} there is an example of the effect on the beam size. 

\begin{figure}[ht]
  \centering
  \includegraphics[width=50mm]{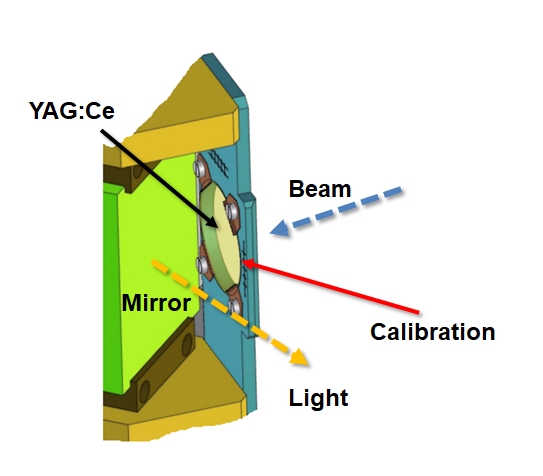}
  \caption{Example of a mounting to avoid the blurring effect. A calibration screen is also hosted in the frame.}
  \label{mounting}
\end{figure}

From the point of view of the observer, being the light emitted all along the crystal, the apparent beam size is much larger with respect to the real one. 
Being this problem produced by the angle between the beam propagation and the light emission, to avoid it is better to place the screen at 90 degrees with respect to the beam axis. 
In Fig.\ref{mounting} there is an example of a scintillator screen mounted with a mirror at 45 degrees on the back. 
In this case the light is emitted in the direction of the beam line and is reflected later by a mirror.

The light emitted is collected, through a vacuum window, by a CCD camera with proper optics. 
In this way a picture of the beam is recorded. 
It is very important to have a proper calibration between pixels and real dimension. 
The best way is to image a calibration pattern placed in the same plane of the screen. 
In the scintillator holder in Fig.\ref{mounting} there is also a pattern where some calibration marks are reported. 

\begin{figure}[h]
  \centering
  \includegraphics[width=80mm]{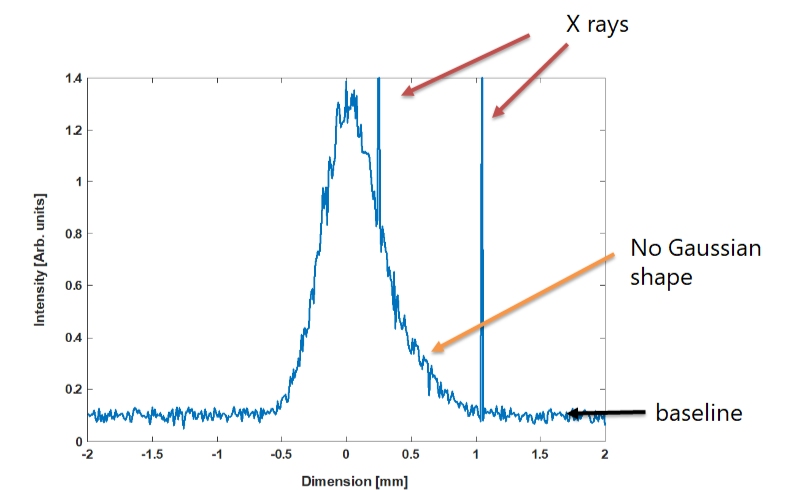}
  \caption{Projection of the beam shape on one of the transverse axis. Some features, typical of a real measurement are reported.}
  \label{rms}
\end{figure}

To retrieve the value of the beam size we have to project the 2D charge distribution on the two vertical and horizontal axes. 
Let us assume a 1D intensity distribution, which can represent the projection of the beam profile on one axis, with $i$ points each of intensity $I{_i}$.
The definition of the second momentum of the distribution is

\begin{equation}
\left\langle {x^2 } \right\rangle  = \frac{{\sum\limits_i {I_i \left( {x_i  - \bar x} \right)^2 } }}{{\sum\limits_i {I_i } }}~.
\label{formula2}
\end{equation}
Be aware that a point with $x = \bar{x}$ gives no contribution, even if its intensity is very large, while a point very far away from $\bar{x}$ gives a huge contribution, even for a small intensity.
In Fig.\ref{rms} there is an example of a beam projection profile, with some typical feature of a real measurement.

A baseline must always be subtracted before starting any calculation, otherwise the points very far from the center can give a huge contribution, completelty spoiling the result of the measurement. 
Also x-rays can give a not negligible contribution, especially when they are far away from the center, where the quadratic dependence of the r.m.s. can amplify their contribution. 
They must be removed.
A simple Gaussian fit in some of the cases is not the best choice, being not compatible with the data. 
These are the reasons why image analysis and data treatment is necessary in order to properly estimate the r.m.s. value of a distribution. 
In \cite{bib:zeuthen}, the reader can find several approaches to data processing in order to discriminate the beam from the surrounding halo.

\subsection{Emittance for space charge dominated beams}

The example on which we focus now is the measurement of the transverse emittance for beam space charge dominated.
It happens always in a photoinjector, just after the gun, where the energy is between few MeV to hundred MeV roughly. 

The envelope equation \cite{bib:reiser} in a drift space is

\begin{equation}
\sigma ''_x  = \frac{{\varepsilon _\mathrm{n}^2 }}{{\gamma ^2 \sigma _x^3 }} + \frac{I}{{\gamma ^3 I_0 \left( {\sigma _x  + \sigma _y } \right)}} \ ,
\label{envelope}
\end{equation}
where $\varepsilon_\mathrm{n}$ is the normalized emittance, $\sigma_x$ and $\sigma_y$ are the beam dimensions, $I$ is the beam current and $I_0$ is the Alfv\'en current.
When the term representing the space charge effect, the second term in this equation, is greater than the emittance term, the beam dynamics is said to be space charge dominated.
Because the space charge contribution decreases as $\gamma^2$, this term is more relevant only at lower energies.
However, because it has the effect of an internal pressure inside the beam, as the emittance, it must be kept negligible to measure the emittance.

\begin{figure}[h]
  \centering
  \includegraphics[width=60mm]{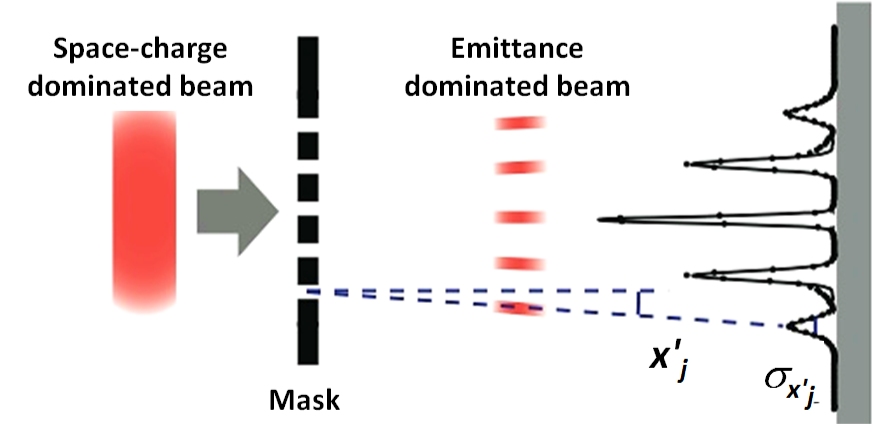}
  \caption{Pepperpot technique: a beam space charge dominated hits a mask. The beamlets emerging from the mask are emittance dominated. After a drift they are imaged on a screen.}
  \label{pepperpot}
\end{figure}

The most common technique is the pepperpot technique \cite{bib:pepperpot}. 
The principle is shown in \Fref{pepperpot}. 
The beam is stopped or heavily scattered by a mask of a material of high atomic number, usually tungsten, while some parts of the beam, called beamlets, pass through small holes or apertures.
Since the charge is significantly suppressed, the contribution of the space charge is negligible and the beamlets are emittance dominated. 
After a proper drift, a scintillator screen produces an image of the beamlets. The relative intensity of each beamlet image is a direct measurement of the transverse beam distribution, $\left\langle {x^2 } \right\rangle$.
In fact, it is a measurement of how much charge is entering any single hole; if the mask geometry is well known, it is a sampling of the beam transverse charge distribution.
The width of each single spot gives a measurement of the beam divergence $\left\langle {x'^2 } \right\rangle$.
A perfectly parallel beam produces images with dimensions equal to the hole size. 
The increase of this dimension is only due to the beam angular divergence, when the space charge contribution is negligible.
Finally the mapping of the beam divergence in different transverse positions allows the reconstruction of the correlation term between position and divergence,$ \left\langle {xx'} \right\rangle$.
In this way, we can measure the second term in \Eref{normemittance}, \ie the geometrical emittance.

In implementing this diagnostic technique, there are some constrains to keep in mind.
First of all, the hole dimensions must be large enough to allow some beam to pass through, to improve the signal-to-noise ratio, but at the same time they must ensure that the beamlets are emittance and not space charge dominated. The ratio between the two terms in the envelope equation is

\begin{equation}
R_0  = \frac{{I\sigma _x^2 }}{{2\gamma I_0 \varepsilon _\mathrm{n}^2 }} \ .
\label{ratio}
\end{equation}
Assuming a uniform beam charge distribution in the holes, the r.m.s. is $\sigma_x={d}/\sqrt {12}$, where $d$ is the diameter of the hole.
To ensure that the beamlets are emittance dominated, $R_0\ll 1$. This constrains the value of $d$. 

The length $L$ of the drift between the mask and the screen is another important parameter, in order to give to the beamlets enough space to develop a dimension greater than the hole size.
The beam size at a distance $L$ from the mask is the square of the quadratic sum of the dimension of the hole (in our case we assumed ${d}/\sqrt {12}$) and the contribution of the beam divergence multiply by the drift length:

\begin{equation}
\sigma _x  = \sqrt {\left( {L \cdot \sigma '_x } \right)^2  + \frac{{d^2 }}{{12}}} \ .
\label{sigma}
\end{equation}
Since $\sigma '_x $ is the parameter to measure, the first term in the square root must be much greater than the second one, setting the limit of the shorter acceptable drift $L$.
In addition, the thickness $l$ of the mask material is very important.
It must be large enough to stop or heavily scatter the beam at a large angle (a critical issue at high energies), but a large thickness can limit the angular acceptance of the hole (resulting in an underestimation of the emittance), which cannot be smaller than the expected angular divergence of the beam, \ie $l < d/2\sigma '_x$.

The better is the sampling of the beam, the better will be the measurement accuracy. 
Mechanical machining of the slits is limited to creating holes of tens or few hundreds of micrometers, while laser techniques can produce definitely better results. 
There are several examples of particular and interesting geometries already realized. 
In \cite{bib:pp1} the authors used a laserbeam machined tungsten disk of 200 $\mu$m thickness, 20 $\mu$m diameter holes and separated by 250 $\mu$m in both dimensions. 
In \cite{bib:pp2} 3 $\mu$m hole size spaced by 50 $\mu$m have been produced. 
The limit of having such small holes and so closed displaced is in the number of the charges that can go through (so in the Signal to Noise Ratio) and in the possible overlap (to avoid!) of the distributions coming from adjacent holes.

It is worth mentioning that this technique is a single shot measurement and not only enables the emittance and the Twiss parameters to be measured but can also be used to reconstruct the entire trace space.
The holes are sampling properties of such a space in different transverse positions.
Every time there is a hole we can measure there the charge in this transverse position and the divergence of the beam. 
The reader can find in \cite{bib:zhang} more details and the formulas to reconstruct the beam parameters starting from the measured quantities.
\Fref[b]{comparison} shows a comparison between simulated and measured trace spaces \cite{bib:cianchi}, demonstrating the effectiveness of this system.

\begin{figure}[ht]
  \centering
  \includegraphics[width=70mm]{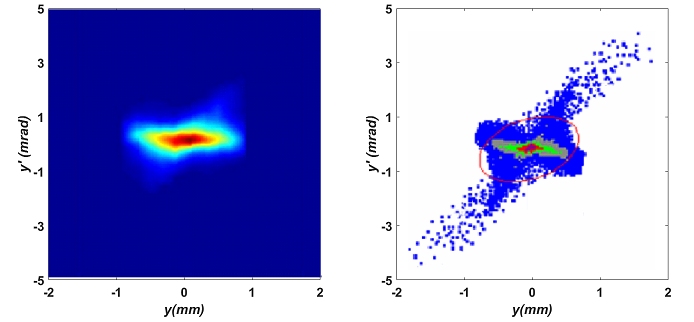}
  \caption{Comparison between measured (left) and simulated (right) trace space \cite{bib:cianchi}}
  \label{comparison}
\end{figure}

The simple device has been also implemented in a more complex instrument \cite{bib:cianchi}.
A long bellow and a motorized screen holder allow to place the mask and the scintillator screen at different longitudinal positions.
In this way the emittance measurement can be done in different longitudinal places. 

\begin{figure}[ht]
  \centering
  \includegraphics[width=90mm]{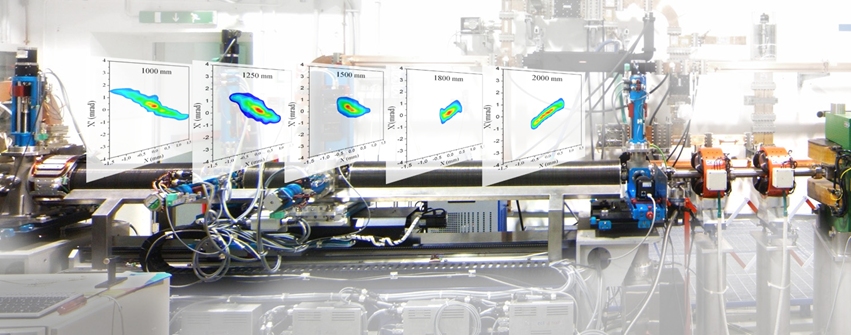}
  \caption{Beam trace space evolution in a RF photoinjector, measured by moving both the mask and the screen together in a range of about 1m downstream a RF gun \cite{bib:cianchi}.}
  \label{emittancemeter}
\end{figure}

In Fig.\ref{emittancemeter} is shown the measurement of the trace space downstream a RF gun.
It is clearly visible the evolution of the trace space. 
This system is very useful to study the beam dynamics in a RF photoinjector. 

\section{Longitudinal diagnostics}

In linacs, especially if they drive FELs, the longitudinal properties are very important. 
It happens often that the right parameters to drive the FEL are not reached along the whole bunch but only in some longitudinal positions (slice).
Also, the peak current, i.e. the charge divided by the bunch length, is a parameter of paramount important in a FEL, and so the correct evaluation of the beam length is fundamental.
Among several methods used for such measurements, the reader can find more details in \cite{bib:cianchicas}, we focus only on two of them, widely used and completely different one to the other. 
TDS (Transverse deflecting structures) are intercepting diagnostics with quasi-single shot capability, while EOS (Electro-Optical-Sampling) is totally not intercepting and single shot technique.

\subsection{Transverse deflecting structure}
\label{sec:RFD}

Transverse deflecting structures \cite{bib:emma,bib:beh}, sometimes also called RF deflectors, are powerful devices, able to attain resolutions of even a few femtoseconds in the X band \cite{bib:xrfd}.
They are quasi-single shot intercepting devices (for the measurement we only need two shots, but they need to be calibrated in multi-shot scenarios, as we'll see later), with the advantages of an inherent self-calibrating nature, simple implementation, and usage.

\begin{figure}[ht]
  \centering
  \includegraphics[width=100mm]{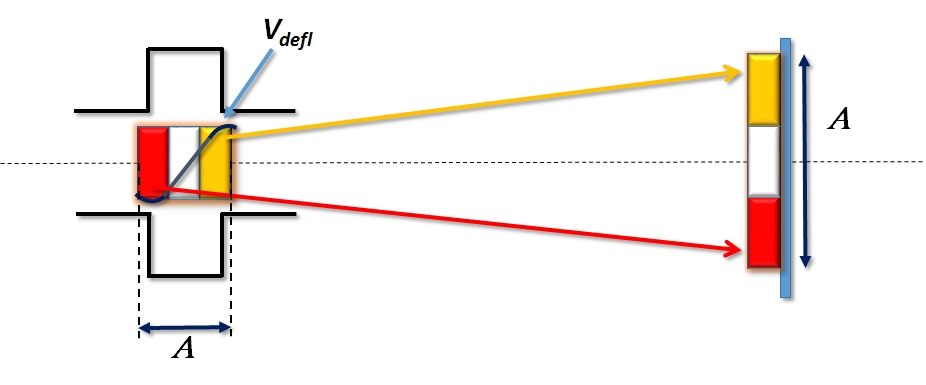}
  \caption{Principle of operation of RF deflector. The longitudinal structure is mapped on the transverse profile because different longitudinal positions experience different transverse kicks.}
  \label{rfd}
\end{figure}

The working principle is shown in \Fref{rfd}.
A time-dependent transverse deflecting voltage is present in a standing or traveling wave structure.
Different parts of the beam, in different longitudinal positions, explore a correlated force that imprints a transverse momentum on the bunch.
After a drift, the imaging on a screen returns the longitudinal charge distribution.
The change in the transverse momentum is $\Delta y'$ = $qV/{pc}$, where $V$ is the deflecting voltage, $q$ is the charge, $p$ is the momentum, and $c$ is the speed of light.
Assuming that the deflecting voltage is of the form $V = V_0 \sin (kz + \varphi )$, where
$k = 2\pi/\lambda $ and $\lambda$ is the RF wavelength, and considering that the bunch length is usually much smaller than such a wavelength (\ie $kz \ll 1$), it is possible to expand the expression
\[\sin (kz + \varphi ) = \sin (kz)\cos (\varphi ) + \cos (kz)\sin (\varphi ) \simeq kz\cos (\varphi ) + \sin (\varphi ) \ .\]
The expression of the change in the transverse momentum given by the structure is
\begin{equation}
\Delta y' = \frac{{qV_0 }}{{pc}}\left[ {kz\cos (\varphi ) + \sin (\varphi )} \right] \ .
\label{deltaxprimo}
\end{equation}

Using a phase value like $\varphi=0$ or $\varphi=\pi$ nulls the second term, and the kick has a simple proportionality to the longitudinal position of the charge.
Using a general expression of the transport matrix in terms of the Twiss parameter (see \cite{bib:rossbach}) the value of the vertical displacement as a function of the longitudinal position is
\begin{equation}
y(z) = y_0  + \left( {\sqrt {\beta \beta _0 } \sin \Delta } \right)y_0 ^\prime   \pm \frac{{qV_0 }}{{pc}}kz\left( {\sqrt {\beta \beta _0 } \sin \Delta } \right)
\label{deflection}
\end{equation}
where $\beta_0$ and $\beta$ are the Twiss parameters on the RF deflector and on the screen, respectively, y$_0$ is the displacement of the charge respect to the axis of the cavity, and $\Delta$ is the betatron phase advance between the RF deflector and the screen.
The $\pm$ sign depends on which zero of the RF phase is chosen, $\sin (\varphi )=0$ for both $\varphi=0$ or $\varphi=\pi$.
The second momentum of the distribution (the quantity that we measure) on the screen is given by
\begin{equation}
\begin{array}{l}
 \left\langle {\left( {y - \left\langle y \right\rangle } \right)^2 } \right\rangle ^ \pm   = \left\langle {y_0^2 } \right\rangle  + \beta \beta _0 \sin ^2 \Delta \left\langle {y_0 {^\prime{^2} }}\right\rangle  - \left\langle {y_0 } \right\rangle ^2  + \beta \beta _0 \sin ^2 \Delta \left( {\frac{{qV_0 }}{{pc}}k} \right)^2 \left\langle {z^2 } \right\rangle +\\
   - \beta \beta _0 \sin ^2 \Delta \left\langle {y_0 ^\prime  } \right\rangle ^2  - \beta \beta _0 \left( {\frac{{qV_0 }}{{pc}}k} \right)^2 \sin ^2 \Delta \left\langle z \right\rangle ^2 +\\ 
  + 2\sqrt {\beta \beta _0 } \sin \Delta \left\langle {y_0 y_0 ^\prime  } \right\rangle  \pm 2\sqrt {\beta \beta _0 } \sin \Delta \left( {\frac{{qV_0 }}{{pc}}k} \right)\left\langle {y_0 z} \right\rangle  - 2\sqrt {\beta \beta _0 } \sin \Delta \left\langle {y_0 } \right\rangle \left\langle {y_0 ^\prime  } \right\rangle +\\ \mp 2\sqrt {\beta \beta _0 } \sin \Delta \left( {\frac{{qV_0 }}{{pc}}k} \right)\left\langle {y_0 } \right\rangle \left\langle z \right\rangle  
  \pm 2\beta \beta _0 \sin ^2 \Delta \left( {\frac{{qV_0 }}{{pc}}k} \right)\left\langle {y_0 ^\prime  z} \right\rangle +\\ \mp 2\beta \beta _0 \sin ^2 \Delta \left( {\frac{{qV_0 }}{{pc}}k} \right)\left\langle {y_0 ^\prime  } \right\rangle \left\langle z \right\rangle  \\ 
 \end{array}
 \label{momento_secondo} 
\end{equation}

In this treatment the beam is assumed monochromatic. 
However, in the presence of a not negligible energy chirp, more terms appear, as reported in \cite{bib:sabato}.

Usually $\left\langle y_0 \right\rangle =\left\langle z \right\rangle  = \left\langle {y_0 ^\prime  } \right\rangle  = 0$.

Let us focus now on this term: 

\begin{equation}
\begin{array}{l}
 \pm 2\sqrt {\beta \beta _0 } \sin \Delta \left( {\frac{{qV_0 }}{{pc}}k} \right)\left\langle {y_0 z} \right\rangle
 \end{array}
\end{equation}

Which is the physical meaning? 
It represents the correlation between the longitudinal position and vertical displacement.

\begin{figure}[ht]
  \centering
  \includegraphics[width=50mm]{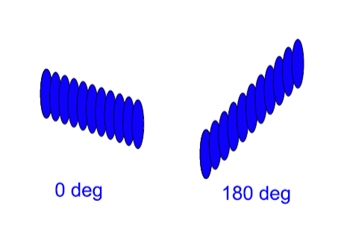}
  \caption{Correlation between longitudinal position and vertical displacement. When the phase of the RFD is switched by 180 degrees, the projections on the vertical axis are different.}
  \label{yz}
\end{figure}

In Fig.\ref{yz} is shown a beam with a correlation between z and y. 
It means that the beam enters the TDS with an angle.
Let us consider for instance that the voltage deflects the beam towards the bottom of the page, clockwise rotation. 
And now, changing the phase for 180 degrees, let us consider a deflection to the top of the page, counterclockwise rotation. 
If you project these two distributions along the vertical axis, as it happens during a measurement, you can find two different projections, see Fig.\ref{yz} and make the vertical projection.
So the result of your measurement is depending on the chosen phase in the TDS, while you expect to have only the dependence on the bunch length. 
If you have a look at \Eref{momento_secondo} this term appears with both sign, depending on the sign of the phase. 
Acquiring two different measurements with opposite phases and averaging the results allow to cancel this term.

It is the same for the term:

\begin{equation}
\begin{array}{l}
  \pm 2\beta \beta _0 \sin ^2 \Delta \left( {\frac{{qV_0 }}{{pc}}k} \right)\left\langle {y_0 ^\prime  z} \right\rangle
 \end{array}
\end{equation}

\begin{figure}[ht]
  \centering
  \includegraphics[width=50mm]{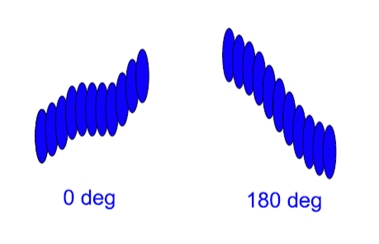}
  \caption{Correlation between longitudinal position and vertical divergence. When the phase of the TDS is switched by 180 degrees, the projections on the vertical axis are different.}
  \label{ypz}
\end{figure}

The physical meaning can be understood looking at the Fig.\ref{ypz}.

In this case the different parts of the beam, in different longitudinal positions, have their own transverse momentum. 
As in the previous case we need to measure at different phases and average the result to remove this term contribution.

When there is no power in the TDS, and using the former conditions, we get

\begin{equation}
\sigma _0^2  = \left\langle {y_0^2 } \right\rangle  + \beta \beta _0 \sin ^2 \Delta \left\langle {y_0 ^{\prime 2 }} \right\rangle  + 2\sqrt {\beta \beta _0 } \sin \Delta \left\langle {y_0 y_0 ^\prime  } \right\rangle 
\label{sigmazero} 
\end{equation}

It represents the dimension of the beam on the screen without any power in the TDS. 
Replacing $\sigma$$_{0}^{2}$ in \ref{momento_secondo}, the measured length is:

\begin{equation}
\sigma  = \sqrt {\sigma _0^2  + \sigma _z^2 } 
\label{RFD_measure} 
\end{equation}

where $\sigma _z^2  = \beta \beta _0 \sin ^2 \Delta \left( {\frac{{qV_0 }}{{pc}}k} \right)^2 \left\langle {z^2 } \right\rangle$.

This term contains the second momentum of the longitudinal charge distribution, i.e. $\left\langle {z^2 } \right\rangle$ the quantity that we are looking for, multiplied by a factor, that we call calibration and we will focus on it later. 

Equating the two terms under square root in Eq.\ref{RFD_measure} is possible to find the resolution:

\begin{equation}
\sigma _z^{res} = \frac{E}{q}\frac{{{\sigma _0}}}{{{V_0}L}}\frac{\lambda }{{2\pi }}
\label{resolution} 
\end{equation}

To increase the resolution of the device the term $\sigma _0^2 $ must be much smaller than $\sigma _z^2$, so the spot on the screen with TDS off must be the smallest achievable. 
Looking at formula \Eref{resolution}, we can see that other ways to improve the resolution are to increase the deflecting voltage (it is the value of the integrated voltage along the structure so, to some extent, increasing the device length can also work) or to decrease the RF wavelength.

\begin{figure}[ht]
  \centering
  \includegraphics[width=50mm]{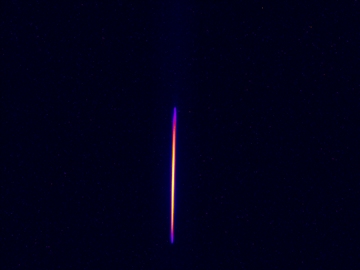}
  \caption{Typical image of a beam streaked by a TDS.}
  \label{typical}
\end{figure}

In a real measurement what we measure is only the length of a vertical (or horizontal) line on a screen $\sigma$, as it is shown in Fig.\ref{typical}.

\begin{figure}[ht]
  \centering
  \includegraphics[width=120mm]{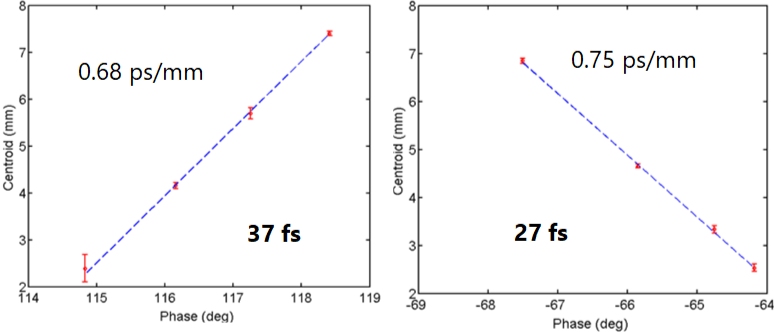}
  \caption{Calibration of the device in the two opposite phases.}
  \label{calibration}
\end{figure}

To convert the dimension that we estimate in pixels (or in $\mu$m) to a real bunch length we need to know the conversion factor between pixels and bunch length, what we have already called calibration. 
Changing the phase of a small amount in the TDS produces the kick of the center of mass. 
Because we know for what amount we modified the phase (and so the time) we can transform the movement in space in time. 
In Fig.\ref{calibration} it is reported the calibration result for two opposite phases.
On the horizontal axis there is the phase and in vertical the relative displacement of the bunch centroid. 

The reader can easily find that the angular coefficient of the lines is the calibration factor that we were looking for. 
At the same time, as expected, the values of the bunch length are different, due to the correlation terms between $z$,$y$ and $y^{\prime}$ that we have already discussed. 
However, the average of the two measurements can give us the correct value of the bunch length.

\subsection{Longitudinal phase space}

\begin{figure}[ht]
  \centering
  \includegraphics[width=50mm]{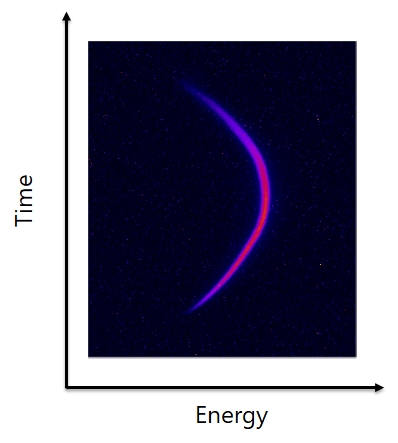}
  \caption{Longitudinal phase space.}
  \label{LPS}
\end{figure}

One of the most interesting feature of a transverse deflecting structure is the possibility to use together with other elements. 
For instance, if we place a dipole, i.e. a device dispersing the beam in energy, after a TDS, i.e. a device dispersing the beam in time, we can have in the single shot the picture of the longitudinal phase space \cite{bib:alesini}, as it is shown in Fig.\ref{LPS}.

\begin{figure}[ht]
  \centering
  \includegraphics[width=40mm]{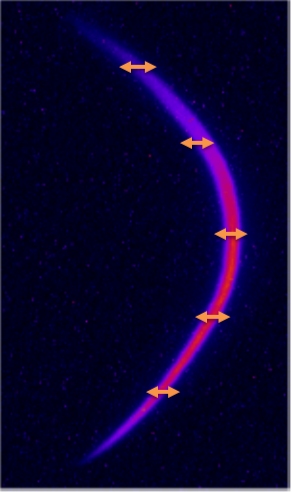}
  \caption{Longitudinal phase space. The arrows represent the contribution of the slice energy spread.}
  \label{LPS_energyspread}
\end{figure}

The projection on the horizontal axis gives the energy spread, while the projection on the vertical axis the bunch length.
This value is very important especially in the FEL. 
How can we extract the slice energy spread from a longitudinal phase measurement?
Which is the effect of the slice energy spread?
The projection in the horizontal axis in Fig.\ref{LPS} gives only the overall energy spread. 
The slice energy spread is highlighted in Fig.\ref{LPS_energyspread}.

The thickness of the distribution is a measurement of the slice energy spread. 
So, the measure of the size of the longitudinal phase space in different vertical position returns the energy spread in different longitudinal positions (slice).
Another feature that is coming for free is the shape itself of the longitudinal phase space. 
In Fig.\ref{LPS_energyspread} is clearly visible a C-like form, where the C shape is perfectly symmetric. 
If you reflect the image on a horizontal axis, up and bottom parts are about the same. 
This is a clear signature of a beam on crest in the RF accelerating structure. 
A beam off-crest would present not anymore this shape, but the C would be rotated. 
Also the picture of the longitudinal phase space allows a simple comparison with simulations tools. 
These are just examples of additional information that can be extracted from such kind of measurements.

The TM$_{11}$-like deflecting modes have a non-zero derivative of the longitudinal electric field on axis. 
This is a general property of the deflecting modes because the deflecting voltage is directly related to the longitudinal electric field gradient through the Panofsky-Wenzel theorem \cite{bib:PWT}.

The transverse accelerating field in the cavity increases the beam slice energy spread.
The energy spread growth is due to the longitudinal electric field that varies linearly with transverse distance.
The induced relative slice energy spread by the TDS is \cite{bib:energyspread}

\begin{equation}
{\sigma _{E/{E_0}}} = \frac{{2\pi }}{\lambda }\frac{{q{V_0}}}{{pc}}{\sigma _{beam}}
\label{induced} 
\end{equation}

where $\lambda$ is the wavelength of the RF, $q$ is the charge, $V_0$ is the deflecting voltage, $p$ is the momentum and $\sigma _{beam}$ is the dimension of the beam at the TDS entrance.
When the resolution of the device is pushed to fs scale, using high voltage and low wavelength, the measurement of the slice energy spread can be completely messed by the induced value in the TDS. 

\subsection{Electro Optical Sampling}

The electro-optic diagnostics technique is a non-destructive and single-shot method.

Large electric fields (of the order of megavolts per metre) applied to optically active crystals
lead to the linear electro-optic effect (the Pockels effect), which is the basis of electro-optic
detection techniques.
In an isotropic medium, the polarization induced by an electric field is always parallel and linear with the electric field vector and related to the field by a scalar factor, the susceptibility.
If the medium is anisotropic, the induced polarization is still linear but not necessarily parallel to the electric field, so the susceptibility is a tensor.
In such a case, the crystal becomes birefringent, with different refraction indices along its principal axes.

In the Electro Optical Sampling diagnostics (EOS) the Coulomb electric field co-propagating with the relativistic electron bunch induces a time-dependent birefringence in an optically active crystal.
A near infrared laser, used as a probe, propagates along the crystal. 

\Fref[b]{index ellipse} shows the so-called ellipse index.

\begin{figure}[ht]
  \centering
  \includegraphics[width=80mm]{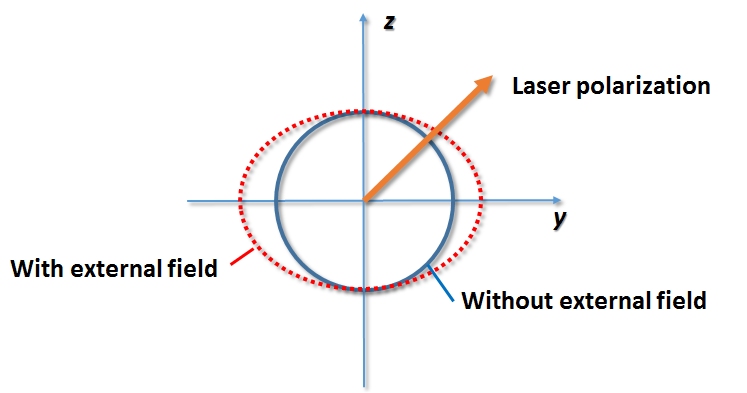}
  \caption{Index ellipse for electro optic crystal. Without external field the index of refraction is the same in every direction. It is not true anymore with an external field.}
  \label{index ellipse}
\end{figure}

Without any external applied field, the crystal is isotropic and the refraction indices are equal along both axes.
However, when the field is applied, the ellipse is distorted and the indices are different.
The polarization of the external laser is in between these axes. 
Owing to the different refraction indices, the propagation speeds along these different axes (c/n$_ {y,z}$ where c is the speed of the light and n$_ {y,z}$ are two different refractive indexes along the axes) are different, and this results in a rotation in the laser polarization.
If we place an orthogonal polarizer before the crystal and another after it, we can observe the laser transmission only when there is an applied electric field. 
So the rotation angle encodes information about the field strength, which reflects the amount of the charge in the beam bunch that produces such a field.
Being E$_L$ the electric field of the laser, the measured electric field is 

\begin{equation}
{E_M}(t) = {E_L}(t){e^{i\Gamma (t)}} 
\label{EOS1} 
\end{equation}

where $\Gamma$ is a factor defined \cite{bib:casalbuoni} as follow:

\begin{equation}
\Gamma (\omega ) = \frac{{2\pi }}{\lambda }L\delta n(\omega ){T_{crystal}}(\omega)
\label{gamma} 
\end{equation}

where L is the crystal length, $\delta$n is the variation in the index of refraction, and T$_{crystal}$ is defined as \cite{bib:van_tilborg} and it is mainly dependent on the real and imaginary index of refraction of the EO crystal in the THz regime.
The rotation of the polarization is directly proportional to the length of the crystal and to the change in the index of refraction. 
This change is proportional to the r$_{41}$ (electro optical coefficient) and to the electric field (E$_{THz}$) travelling with the charge.

\begin{equation}
\delta n \propto \frac{{n_0^3}}{4}{r_{41}}{E_{THz}}
\label{EOS3} 
\end{equation}

So the overall polarization rotation is proportional to the electric field of the moving charge. 
If we consider that E$_{THz}$ can be produced by a bunch of particles, the measurement of $\Gamma$ can be a measurement of the charge distribution of the incoming beam.

The value of r$_{41}$  is 4.25 10$^{-12}$ m/V for ZnTe and 1.0 10$^{-12}$ m/V for GaP, two popular crystals used for this kind of measurement. 
Being the electro-optical coefficient so small, from \Eref{gamma} one might think that an increase in the crystal length can increase the effect. 
It is true, but this does not help in the measurements. 
Indeed it can be a disadvantage.
First of all the E$_{THz}$ pulse and the laser pulse propagates at different speed inside the crystal, so there is slippage in phase that increases with the crystal length. 
Any crystal is also limited by the absorption band. 
Frequency higher than this cutoff cannot propagate. 
It is about 5 THz for ZnTe and 10 THz for GaP. 
The transmission is also not linear at all the frequencies and so the E$_{THz}$ can suffer a dispersion inside the crystal. 

\begin{figure}[ht]
  \centering
  \includegraphics[width=160mm]{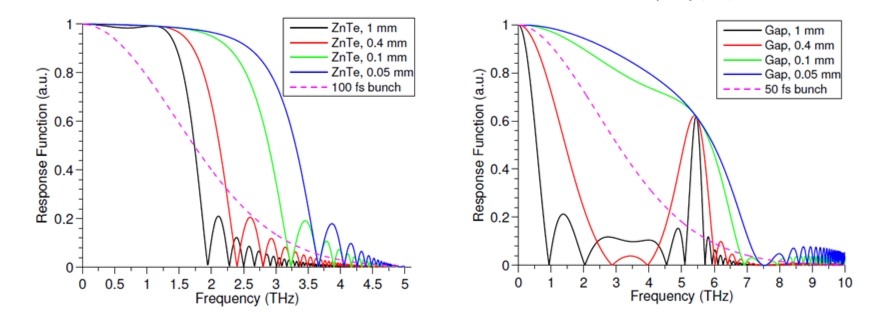}
  \caption{left: EO response function of ZnTe for crystal thickness of 50 $\mu$m, 100 $\mu$m, 400 $\mu$m and 1 mm. The power spectrum of a 100 fs electron bunch is showed for comparison.
Right: EO response function of GaP for crystal thickness of 50 $\mu$m, 100 $\mu$m, 400 $\mu$m and 1 mm. The power spectrum of a 50 fs electron bunch is showed for comparison \cite{bib:pompili}.}
  \label{response_function}
\end{figure}

In the \Fref{response_function} there are the plots of the response function of ZnTe e GaP, taking into account slippage and frequency dispersion.
In the plots it is superimposed also the shape of the frequency spectrum of a bunch of 100 fs and 50 fs respectively. 
It is clear that for the case of 50 fs for instance, the crystal thickness cannot be larger than 100 $\mu$m otherwise there are zeros and oscillations in the frequency range of the spectrum that we want to reconstruct. 

In the definition of the parameters of the system to build, the following considerations have to be made.
The first definition is the range of the bunch length that we want to measure. 
This parameter is fundamental in the choice of the crystal (the frequency content of the bunch spectrum cannot arrive to the absorption band of the crystal), the readout scheme (some schemes are inherently limited in time resolution) and the probe laser that must be shorter than the bunch length.
Later must be considered the electro-optical coefficient of the crystal to evaluate if the signal is detectable, i.e. if there is a rotation of the polarization.
Finally the thickness of the crystal must be chosen as a compromise between the need to have a signal, from \Eref{gamma}, and the request to preserve the spectrum, as shown in Fig.\ref{response_function}.

So far we said that an electro optical effect, induced by a charged beam, can rotate the polarization of the radiation of a probe laser, and it can do of a certain amount proportional to the longitudinal beam profile.
We have seen that the effect is very small and the thickness and the type of the crystal play a role.
Now it is time to understand how to retrieve the information of longitudinal profile from the signal. 

\begin{figure}[ht]
  \centering
  \includegraphics[width=120mm]{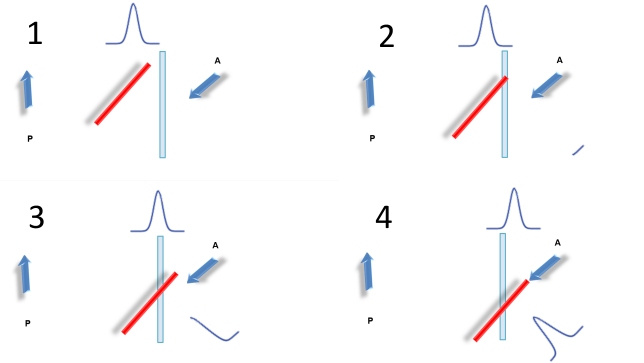}
  \caption{EOS spectral decoding. P and A are two orthogonal polarizers, the EOS crystal is the vertical rectangle, the laser arrives with an angle and is the red rectangle. In the upper part of the figure there is the electric field intensity of the beam propagating close to the crystal.}
  \label{EOS_scheme}
\end{figure}

There are several readout schemes. 
The resolution achievable is different changing the scheme, and the implementation is also varying. 
Again, we are not going to review all the different schemes of EOS. 
The reader can find details about the spectral decoding in \cite{bib:wilke} and about the temporal decoding in \cite{bib:eos}.
Here we focus on spatial decoding \cite{bib:cavalieri}.
To understand the principle of the method we can refer to Fig.\ref{EOS_scheme}.

There are two orthogonal polarizers, called P and A.
The crystal is the vertical rectangle, while the laser is represented by the red inclined rectangle.
The laser arrives with an angle on the crystal. 
Without the coulomb field of the beam there is not a polarization rotation and no signal is transmitted. 
In the frame called \textbf{1} both laser and coulomb field are not on the crystal and the transmission is zero. 
In frame number \textbf{2} the tail of the bunch, and so a weak electric field, is close to the crystal. 
There is a rotation of the polarization of the laser, but the effect is weak, reflecting the intensity of the Coulomb field. 
 In frame number \textbf{3} the center of the bunch is now transiting close of the crystal. 
The electric field is strong and so also the laser rotation.
It reflects in a large signal on the detector.
Finally in frame number \textbf{4} again only the tail of the bunch is now present close to the crystal, the Coulomb field is weak as well as the laser rotation and the signal intensity. 
The key feature of this scheme is the angle between laser and the crystal. 
It creates a correlation between time and position. 
Different parts of the laser cross the crystal in different time, as it is shown in Fig.\ref{EOS_scheme}.
Small angles are required to achieve high temporal resolution by improving both the velocity matching between the laser and the THz pulse inside the crystal, while larger angles increase the temporal window, simplifying the synchronization between the laser and the electron beam.
 
\begin{figure}[ht]
  \centering
  \includegraphics[width=50mm]{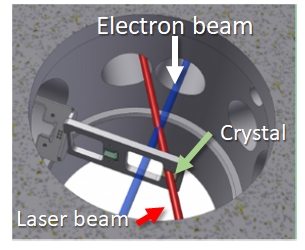}
  \caption{A sketch of the interaction of the laser and the electron beam with the crystal.}
  \label{EOS_drawing}
\end{figure} 
 
A possible mounting scheme is shown in Fig.\ref{EOS_drawing}.
The electron beam is passing close to a crystal, where a laser beam is moving with an angle with respect to the EO crystal.
The signal is then recorded with the help of normal CCD camera. 

The reader can consider that usually the resolution of the EOS in measuring bunch length is not better than 40 fs. 
Compared with TDS, that can arrive close to fs, it seems not so attractive. 
However, the cost of the device can be incredibly different. 
For instance a x-band power station with the TDS and all the ancillaries, like waveguides and chillers can easily arrive to 1M\euro.
The cost of EOS can be extremely variable but it is at least one order of magnitude less. 

The main issue in this setup is the proper synchronization between an electron beam and a probe laser. 
The jitter of the laser with respect to the beam can further improve this problem.
In \cite{bib:veronese} the authors used a flexible but expensive setup with a fiber laser at 78 MHz placed directly close to the vacuum chamber. 
This setup needs an intensified camera to select one single pulse from the 78 MHz train and an external unit to synchronize the camera to the radiofrequency.
However, in \cite{bib:pompili2} the same laser, used for the photoinjector that produces the photoelectrons, is splitted to be used as probe laser in EOS.
In this way it is naturally synchronized with the beam and there is no need neither of an intensified camera as well as the external unit.  
However, the path of the laser and the trajectory of the electron inside the machine are not equal. 
So a delay on the laser line must be introduced. 
A fast photodiode is used to see the laser signal and to adjust the delay in time with a signal coming from the beam, usually a pickup. 
In this way a rough synchronization, in the order of tens of ps is done, while the small adjustment can be done with a remotely controlled delay line looking directly at the EOS signal. 

\begin{figure}[ht]
  \centering
  \includegraphics[width=50mm]{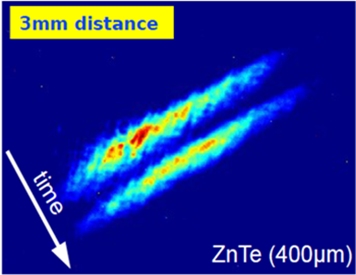}
  \caption{Measurement of EOS with two bunches, separated approximately by 900 fs.}
  \label{EOS_meas}
\end{figure}

From the same paper we can also show an image of the signal of the EOS when there are two bunches, see Fig.\ref{EOS_meas}, using a ZnTe crystal 400 $mu$m thick placed at 3 mm from the beam.

The reader can imagine that the beam is normal to the plane of the foil, as the laser. 
The camera is imaging the crystal surface. 

\begin{figure}[ht]
  \centering
  \includegraphics[width=80mm]{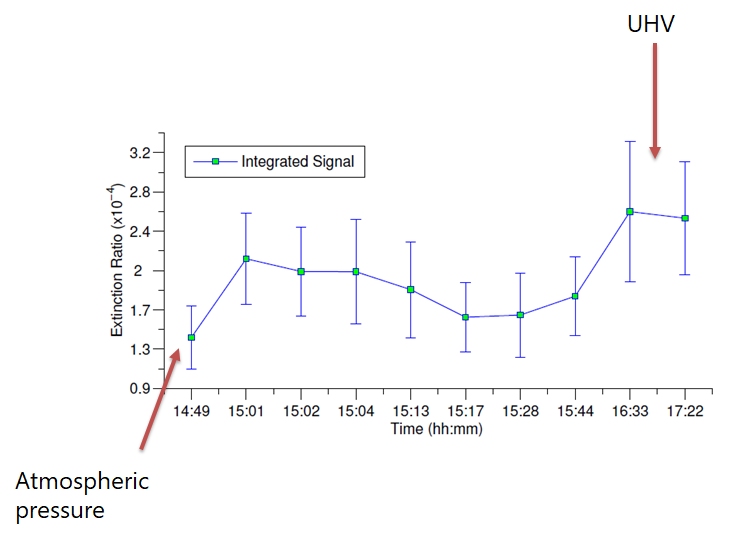}
  \caption{Residual birifringence induced by a mechanical stress on the vacuum windows.}
  \label{vacuum}
\end{figure}

The EOS is a very sophisticated diagnostics and even small effect must be taken into account. 
When we put the two polarizers orthogonal one each other we do not expect to see any signal. 
However, every optical material that is crossed by the laser can have a small residual birefringence that can rotate the polarization. 
In Fig.\ref{vacuum} there is a measurement of the extinction ratio between two orthogonal polarizers, changing the pressure inside a vacuum chamber. 
The laser must cross of course the windows to enter and to leave the chamber. 

The mechanical stress induced by the difference in the pressure between the two sides of the vacuum window produces a small birefringence that decrease the extinction ratio. 
To correct this problem a quarter waveplate is usually introduced in the laser line. 
Without the beam, the plate is rotated until there is not anymore signal transmitted by the polarizers.

\section{Conclusions}

There is always a big gap between the diagnostics on the paper and the practical
realization of a device.
The mechanical drawings, the implementation, the definition of the detectors, the understanding of what you are really measuring
are unavoidable steps in order to build a successful device.

Patience, precision, curiosity are the main qualities that can drive you to the success.

\section{Acknowledgments}

I would like to thank Enrica Chiadroni, Andrea Mostacci and Riccardo Pompili for the useful discussions and the help in revising the document.

\end{document}